\documentclass[reprint,amsmath,amssymb,amsart,prb,aps,superscriptaddress,twocolumn,float]{revtex4-2}

\usepackage{graphics}
\usepackage{graphicx}
\usepackage{epsfig}
\usepackage{multirow}
\usepackage{dcolumn}
\usepackage{xspace}
\usepackage{array}
\usepackage{tabularx}
\usepackage{gensymb}
\usepackage{bigints}
\usepackage{bm}
\usepackage{subfigure}
\usepackage{color}
\usepackage{mathtools} 
\usepackage{amsmath} 
\usepackage{courier} 

\begin{document}


\title{Strong-coupling anisotropic superconductivity in hexagonal HfRuAs from anisotropic Migdal-Eliashberg theory}
\author{P. V. Sreenivasa Reddy}
\email{peramsreenivas@gmail.com}
\affiliation{Department of Physics, National Taiwan University, Taipei 10617, Taiwan\looseness=-1} 

\author{Guang-Yu Guo}
\email{gyguo@phys.ntu.edu.tw}
\affiliation{Department of Physics, National Taiwan University, Taipei 10617, Taiwan\looseness=-1} 
\affiliation{Physics Division, National Center for Theoretical Sciences, Taipei 10617, Taiwan\looseness=-1}

\date{\today}

\begin{abstract}
We present a comprehensive theoretical investigation of the superconducting (SC) properties of hexagonal HfRuAs ($h$-HfRuAs) by solving anisotropic Migdal--Eliashberg (ME) equations with the inputs from \textit{ab initio} calculations of electronic structure, phonon dispersion and electron phonon coupling matrix elements. The calculated Eliashberg spectral function reveals strong electron--phonon coupling (EPC) with a constant $\lambda \approx 1.56$, dominated by low-frequency phonon modes associated primarily with Hf and Ru vibrations. The SC state is characterized by a single anisotropic gap with overall $s$-wave symmetry, as evidenced by the fully gapped quasiparticle density of states. The momentum-resolved EPC and SC gap exhibit pronounced anisotropy across different Fermi surface sheets, with the largest variations occurring on the hole-like bands. The SC gap is centered around $\Delta \approx 2.9$ meV with a spread of $\sim 0.8$ meV, indicating significant multiband anisotropy. The resulting gap ratio $2\Delta(0)/k_B T_c \approx 4.2$ exceeds the BCS weak-coupling limit, establishing $h$-HfRuAs as a strong-coupling superconductor. The calculated transition temperature, $T_c$, agrees in the order of magnitude  with experiments. Overall, our results identify $h$-HfRuAs as a phonon-mediated, strongly coupled anisotropic superconductor and provide detailed insights into the role of momentum-dependent electron--phonon interactions in determining its SC properties.

\end{abstract}


\maketitle


\section{Introduction}

Superconductivity in ternary equiatomic intermetallic compounds of the form TT$^{\prime}$X (where T and T$^{\prime}$ are transition metals and X = Si, P, Ge, As) is commonly observed when T$^{\prime}$ is a noble metal~\cite{Zhong1986, Wang1987, Shirotani1999, Meisner1983-1, Shirotani2003, Ruan2016, Meisner1983-2, Meisner1983-3, Kase2016, Shirotani1997, Shirotani1995, Shirotani2000}. These compounds crystallize predominantly in three structural types: the hexagonal ZrNiAl-type (also known as Fe$_2$P-type) structure (h-phase)~\cite{Shirotani1999,Barz1980}, the orthorhombic TiNiSi-type structure (o-phase)~\cite{Zhong1986}, and the orthorhombic TiFeSi-type structure (o$^{\prime}$-phase)~\cite{Meisner1983-1,Shirotani1997}. Among these, h-phase compounds such as HfRuP and ZrRu(Si, P, As), which exhibit superconducting (SC) transition temperatures ($T_c$) exceeding 10 K~\cite{Shirotani1999,Meisner1983-3,Shirotani2000,Barz1980}, have attracted considerable attention. Furthermore, several members of this family, including HfRuP, HfRuAs, ZrRuAs, and ZrRuP, have been reported to host nontrivial topological electronic states~\cite{Qian2019}. 

Recent magnetization and electrical resistivity measurements for h-phase ZrRuAs reported SC $T_c$ of approximately 6.5 K~\cite{Qian2019} and 7.9 K~\cite{D.Das2021}, which are considerably lower than the earlier experimental range of 10.2--11.9 K~\cite{Meisner1983-3}. Such discrepancies are generally attributed to differences in synthesis conditions, heat-treatment procedures, crystallinity, stoichiometry, and defect concentration. From a theoretical perspective, first-principles calculations for h-phase ZrRuAs also show a wide variation in the predicted $T_c$, ranging from approximately 6.1 to 11.1 K~\cite{Tutuncu2020,Ren2025}. Similar discrepancies between theoretical and experimental $T_c$ values have also been reported in related compounds such as HfRuP and ZrRuP~\cite{Ren2025,Barz1980,Bagci2019,Barz1980}, highlighting the sensitivity of superconductivity in this family to structural details, computational parameters, and electron--phonon coupling (EPC) strength.

Among these compounds, recent experiments on hexagonal HfRuAs ($h$-HfRuAs) reported superconductivity with $T_c$ ranging from 4.3 to 7.25 K~\cite{Meisner1983-1,Zhou2025}. The experimentally estimated EPC constant $\lambda_{ep}\approx0.73$ places $h$-HfRuAs in the moderate-coupling regime, while the reduced specific-heat jump $\Delta C/\gamma T_c\approx1.03$ suggests a weak-coupling fully gapped SC state with $s$-wave pairing symmetry~\cite{Zhou2025}. However, despite the growing experimental interest, a comprehensive theoretical understanding of the SC pairing mechanism, momentum-dependent EPC, and anisotropic SC gap structure in $h$-HfRuAs is still lacking. In particular, fully anisotropic Migdal--Eliashberg (ME) calculations~\cite{Migdal1958,Eliashberg1960,Eliashberg1961} are necessary to resolve the multiband SC properties and Fermi surface (FS) dependent pairing interactions in this material.

In this work, we present a comprehensive first-principles investigation of the SC properties of $h$-HfRuAs within the fully anisotropic ME formalism~\cite{Migdal1958,Eliashberg1960,Eliashberg1961}, using \textit{ab initio} calculations of the electronic structure, phonon dispersion, and electron--phonon coupling matrix elements combined with Wannier interpolation~\cite{Marzari2012,Giustino2007,Giustino2017}.
Our calculations reveal that $h$-HfRuAs is a strongly coupled phonon-mediated superconductor characterized by a single anisotropic SC gap with overall $s$-wave symmetry. We further elucidate the momentum-resolved EPC, SC gap anisotropy, and the dominant phonon modes responsible for superconductivity in this system.

The rest of this paper is organized as follows. In Sec.~II, we briefly summarize the anisotropic ME formalism and the theoretical quantities employed in this study. Section~III presents the computational details of the first-principles calculations. In Sec.~IV, we discuss the crystal structure and phonon properties of $h$-HfRuAs. The electronic structure and FS characteristics are presented in Sec.~V. Section~VI reports the EPC properties and their momentum dependence on the FS. In Sec.~VII, we present the SC properties, including the momentum-dependent SC gap, $T_c$, and quasiparticle density of states. Finally, the main conclusions are summarized in Sec.~VIII.

\section{Methodology}

The technical details of the anisotropic ME formalism~\cite{Migdal1958,Eliashberg1960,Eliashberg1961} implemented in the EPW code have been extensively discussed in Refs.~\cite{Margine2013,Ponce2016}. Here, we briefly summarize the essential theoretical framework adopted in the present calculations.

The superconducting properties, including the momentum-dependent gap, transition temperature ($T_c$), and quasiparticle density of states (DOS), were investigated within the fully anisotropic Migdal--Eliashberg (ME) formalism~\cite{Allen1982,Margine2013}. In conventional phonon-mediated superconductors, Cooper pairing occurs predominantly within a narrow energy window around the Fermi level ($E_F$), typically of the order of the characteristic phonon energy $\hbar\omega_{ph}$. Therefore, only electronic states close to the FS contribute significantly to superconductivity.

Within the anisotropic ME formalism, the SC state is described through two coupled nonlinear equations: one for the mass-renormalization function $Z_{n\mathbf{k}}(i\omega_m)$ and the other for the SC gap function $\Delta_{n\mathbf{k}}(i\omega_m)$. The anisotropic ME equations on the imaginary-frequency axis are given by
\begin{align}
Z_{n\mathbf{k}}(i\omega_m)&=1+\frac{\pi T}{\omega_m}\sum_{n'\mathbf{k}'m'}W_{n'\mathbf{k}'}\frac{\omega_{m'}}{\sqrt{\omega_{m'}^2+\Delta_{n'\mathbf{k}'}^2(i\omega_{m'})}}\nonumber\\
&\qquad \times\lambda(n\mathbf{k},n'\mathbf{k}',m-m'),
\end{align}
and
\begin{align}
Z_{n\mathbf{k}}(i\omega_m)\Delta_{n\mathbf{k}}(i\omega_m)&=\pi T\sum_{n'\mathbf{k}'m'}W_{n'\mathbf{k}'}\nonumber\\
&\times\frac{\Delta_{n'\mathbf{k}'}(i\omega_{m'})}{\sqrt{\omega_{m'}^2+\Delta_{n'\mathbf{k}'}^2(i\omega_{m'})}}\nonumber\\
&\times\Big[\lambda(n\mathbf{k},n'\mathbf{k}',m-m')\nonumber\\&\qquad-N_FV(n\mathbf{k}-n'\mathbf{k}')\Big].
\end{align}
Here, $i\omega_m=i(2m+1)\pi T$ are fermionic Matsubara frequencies, where $m$ is an integer and $T$ is the temperature. The quantity $W_{n\mathbf{k}}$ represents the normalized electronic weight at the FS and is written as
$W_{n\mathbf{k}}=\frac{\delta(\varepsilon_{n\mathbf{k}}-E_F)}{N_F}$,
where $N_F$ is the electronic density of states at the $E_F$ and $\delta$ denotes the Dirac delta function. The quantity $V(n\mathbf{k}-n'\mathbf{k}')$ represents the screened Coulomb interaction between electronic states $n\mathbf{k}$ and $n'\mathbf{k}'$. In practical calculations, this interaction is commonly replaced by the empirical Morel--Anderson Coulomb pseudopotential $\mu_c^*$~\cite{Morel1962}.

The anisotropic electron--phonon coupling kernel entering the ME equations is given by 
\begin{align}
\lambda(n\mathbf{k},n'\mathbf{k}',m-m')&=\int_0^{\infty} d\omega\,\frac{2\omega\,\alpha^2F(n\mathbf{k},n'\mathbf{k}',\omega)}{(\omega_m-\omega_{m'})^2+\omega^2}.
\end{align}
The anisotropic Eliashberg spectral function is expressed as
\begin{align}
\alpha^2F(n\mathbf{k},n'\mathbf{k}',\omega)&=N_F\sum_\nu|g_{nn'\nu}(\mathbf{k},\mathbf{q})|^2\nonumber\\&\times\delta(\omega-\omega_{\mathbf{q}\nu}),
\end{align}
where $\mathbf{q}=\mathbf{k}'-\mathbf{k}$ and $\omega_{\mathbf{q}\nu}$ is the phonon frequency for branch $\nu$.

The electron--phonon matrix elements were computed within density-functional perturbation theory (DFPT) as
\begin{equation}
g_{nn'\nu}(\mathbf{k},\mathbf{q})=\frac{1}{\sqrt{2\omega_{\mathbf{q}\nu}}}\left\langle\psi_{n'\mathbf{k+q}}\left|\partial_{\mathbf{q}\nu}V_{sc}\right|\psi_{n\mathbf{k}}\right\rangle ,
\end{equation}
where $\partial_{\mathbf{q}\nu}V_{sc}$ is the derivative of the self-consistent potential associated with a phonon perturbation of wave vector $\mathbf{q}$ and branch index $\nu$. The electron--phonon matrix elements computed on coarse Brillouin zone (BZ) meshes were subsequently interpolated onto dense $\mathbf{k}$- and $\mathbf{q}$-point grids using maximally localized Wannier functions as implemented in the EPW code~\cite{Margine2013,Ponce2016,Lee2023}.

The phonon mode resolved EPC strength $\lambda_{\mathbf{q}\nu}$ is given by
\begin{align}
\lambda_{\mathbf{q}\nu}&=\frac{1}{N_F\omega_{\mathbf{q}\nu}}\sum_{nn'}\int_{BZ}\frac{d\mathbf{k}}{\Omega_{BZ}}|g_{nn'\nu}(\mathbf{k},\mathbf{q})|^2\nonumber\\
&\times\delta(\varepsilon_{n\mathbf{k}}E_F)\delta(\varepsilon_{n'\mathbf{k+q}}-E_F),
\end{align}
where $\Omega_{BZ}$ is the BZ volume.

The momentum-resolved EPC strength is given by  
\begin{equation}
\lambda_{n\mathbf{k}}=\frac{1}{N_F}\sum_{n'\nu\mathbf{q}}\int_{BZ}\frac{d\mathbf{q}}{\Omega_{BZ}}
\frac{|g_{nn'\nu}(\mathbf{k},\mathbf{q})|^2}{\omega_{\mathbf{q}\nu}}
\delta(\varepsilon_{n'\mathbf{k+q}}-E_F).
\end{equation}

The anisotropic ME equations~\cite{Allen1982,Margine2013} were solved self-consistently on the imaginary-frequency axis. The superconducting gap function on the real-frequency axis, required for the DOS calculations, was obtained via analytic continuation of the imaginary-axis solutions.

The SC quasiparticle DOS normalized by the normal-state DOS at $E_F$ was calculated as
\begin{align}
\frac{N_s(\omega)}{N_F}&=\sum_n\int_{BZ}\frac{d\mathbf{k}}{\Omega_{BZ}}\frac{\delta(\varepsilon_{n\mathbf{k}}-E_F)}{N_F}\nonumber\\
&\times\mathrm{Re}\left[\frac{\omega}{\sqrt{\omega^2-\Delta_{n\mathbf{k}}^2(\omega)}}\right].
\end{align}
The normalized quasiparticle DOS provides direct information about the SC gap symmetry. A U-shaped low-energy DOS is characteristic of a fully gapped nodeless superconducting state, whereas a V-shaped low-energy DOS indicates nodal superconductivity.

\section{computational details}

The electronic structure, lattice dynamics, and EPC calculations were performed using the \textsc{Quantum ESPRESSO} package~\cite{Giannozzi2009,Giannozzi2017} within the framework of density functional theory (DFT)~\cite{Perdew1985} and density functional perturbation theory (DFPT)~\cite{Baroni2001}. The Perdew--Burke--Ernzerhof revised for solids (PBEsol) exchange--correlation functional was employed together with optimized norm-conserving Vanderbilt relativistic pseudopotentials~\cite{Hamann2013}. 
A kinetic-energy cutoff of 80 Ry was used for the plane-wave basis expansion. The BZ integrations were carried out using the Methfessel--Paxton smearing scheme with a smearing width of 0.02 Ry. The self-consistent charge density calculations were performed on a $\Gamma$-centered $8 \times 8 \times 12$ {\bf k}-point mesh. For the density of states (DOS) and FS calculations, a denser {\bf k}-point mesh of $32 \times 32 \times 48$ was employed to achieve accurate resolution of the electronic states near the $E_F$.
The phonon dispersion relations and the linear variation of the self-consistent potential were computed within DFPT~\cite{Baroni2001} using a $4 \times 4 \times 6$ {\bf q}-point mesh. The convergence threshold for the self-consistent calculations was set to $10^{-14}$ Ry to ensure accurate evaluation of the phonon frequencies and EPC matrix elements.

The SC properties were investigated within the fully anisotropic ME formalism~\cite{Migdal1958,Eliashberg1960,Eliashberg1961} using the EPW code~\cite{Margine2013,Ponce2016,Lee2023,Noffsinger2010}. Since the solutions of the anisotropic ME equations are highly sensitive to the sampling of the electron--phonon matrix elements near the $E_F$, dense {\bf k}- and {\bf q}-point meshes are required for converged calculations. To efficiently achieve this, the electron--phonon matrix elements calculated on coarse BZ meshes were interpolated onto dense meshes using maximally localized Wannier functions, as implemented in EPW~\cite{Margine2013}.
The Wannier interpolation was constructed from first-principles electronic structure calculations performed on a uniform $8 \times 8 \times 12$ {\bf k}-point mesh. The anisotropic ME equations were subsequently solved using fine {\bf k}- and {\bf q}-point meshes of $32 \times 32 \times 48$. A Fermi-window width of 0.5 eV around the $E_F$ was adopted in the anisotorpic ME calculations. The convergence of the SC gap function, momentum-resolved EPC strength, and SC $T_c$ with respect to the fine {\bf k}- and {\bf q}-mesh samplings was carefully checked to ensure the reliability of the calculated SC properties.

\begin{figure}
\includegraphics[width=80mm]{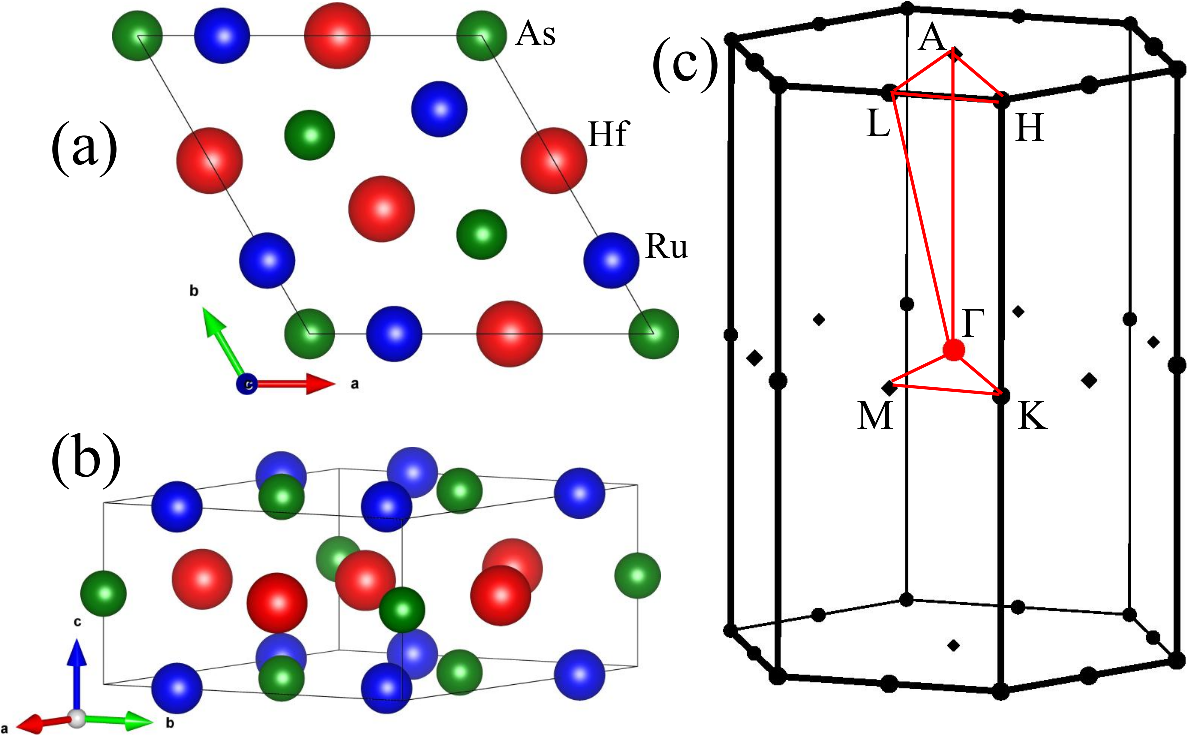}
\caption{(a) A $z$-axis view,  (b) side view of the crystal structure of $h$-HfRuAs. (c) The bulk BZ of $h$-HfRuAs along with high symmetry point labels.}
\end{figure}

\begin{figure}
\includegraphics[width=80mm]{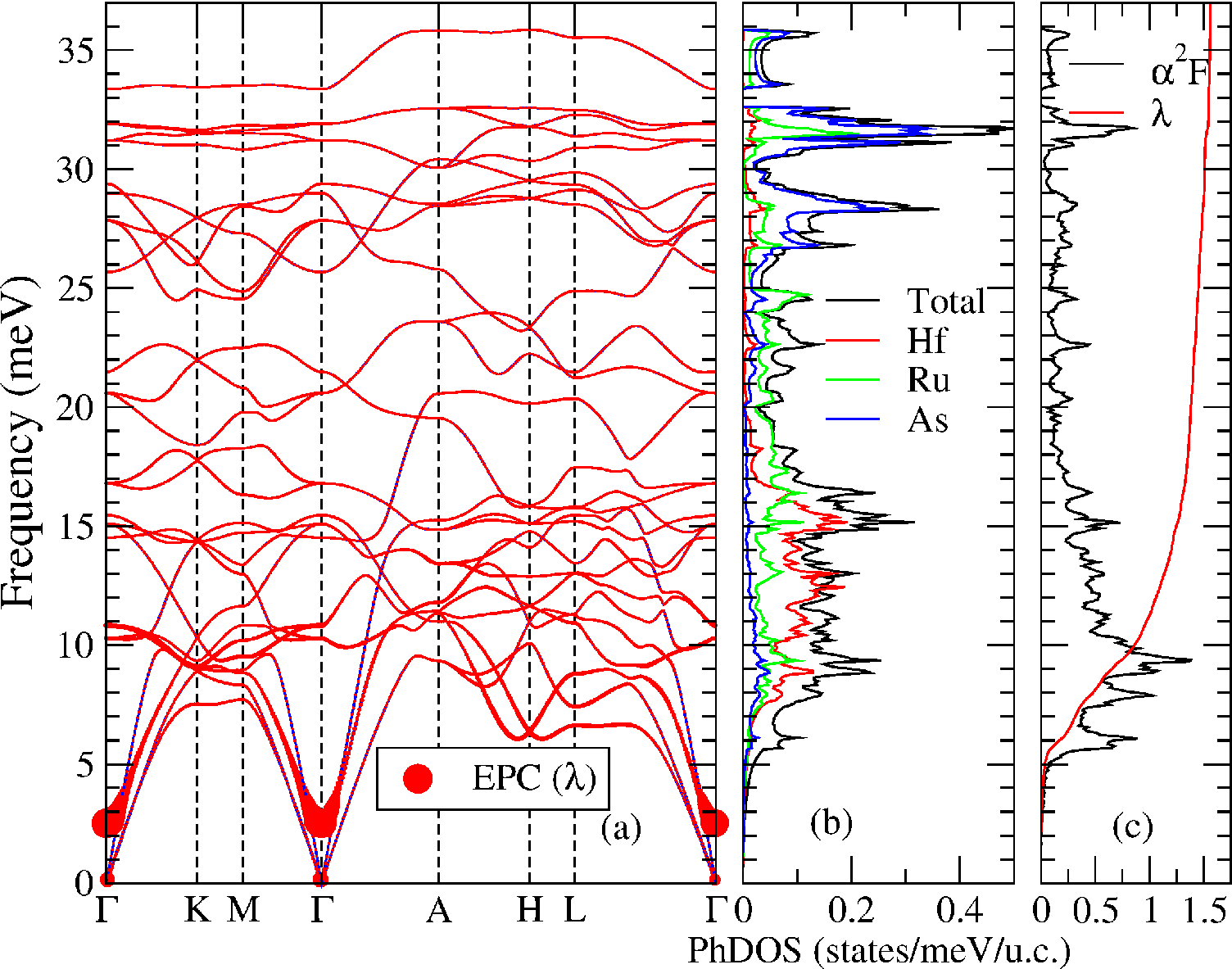}
\caption{(a) Phonon dispersion relations along high-symmetry directions.
Red circle size reflects the EPC strength. (b) Total and atom projected phonon DOS (PhDOS), (c) Eliashberg spectral function ($\alpha^2$F) and accumulative EPC constant ($\lambda(\omega)$) of $h$-HfRuAs.}
\end{figure}

\begin{figure*}
\includegraphics[width=160mm]{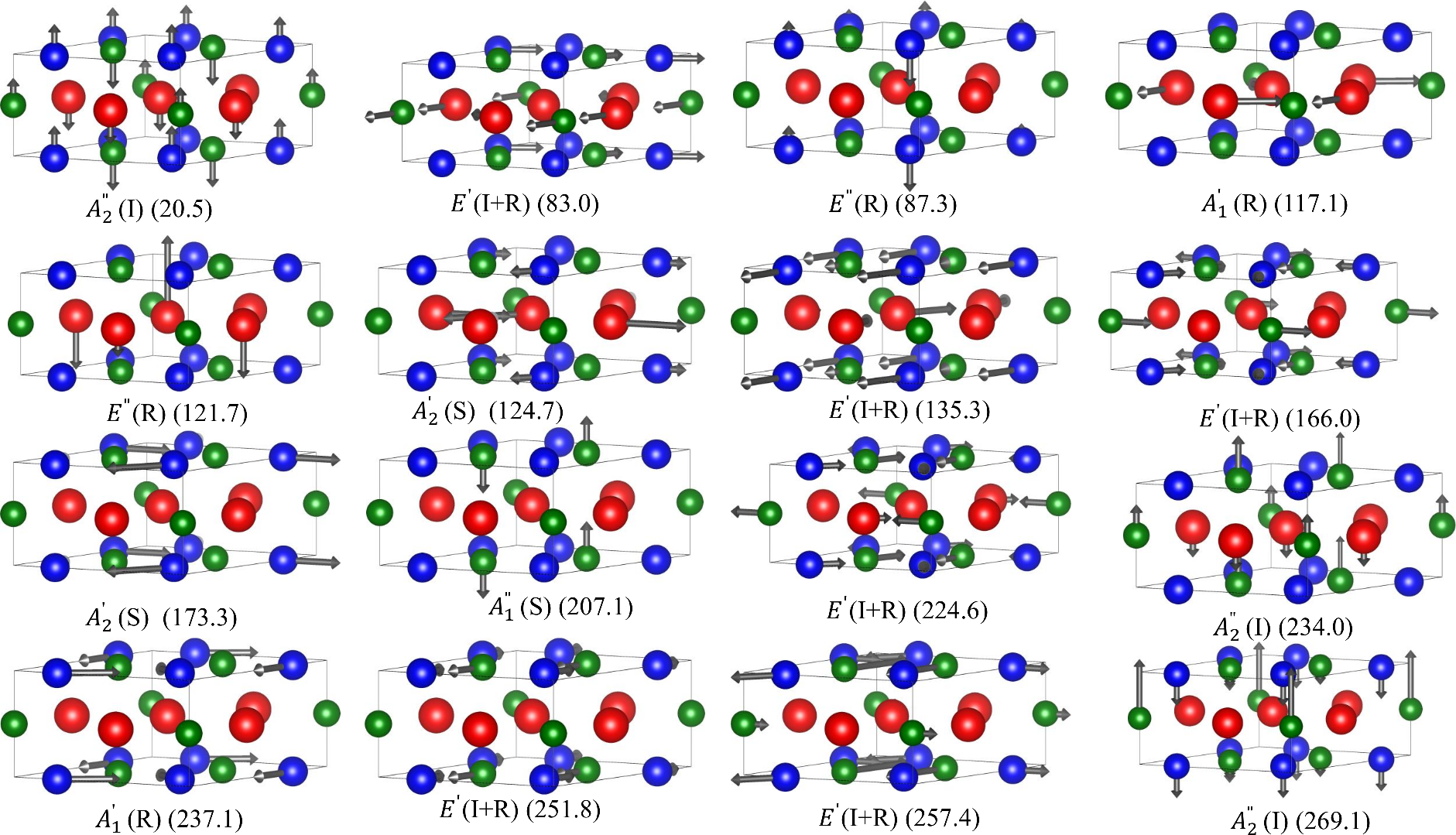}
\caption{Displacement patterns for all the optical phonon modes of $h$-HfRuAs at the $\Gamma$ point. Each panel shows the atomic vibration
pattern associated with a specific irreducible representation. The optical activity of each mode is labeled as infrared active (I), Raman active
(R), or silent (S). The frequencies shown inside the parentheses are in the unit of cm$^{-1}$.}
\end{figure*}

\begin{table}
\caption{Calculated zone-center optical phonon frequencies ($\nu$) and their eigen characters for $h$-HfRuAs. Since experimental Raman spectra are not yet available, the calculated frequencies are reported in units of cm$^{-1}$ to facilitate future experimental comparisons. Infrared-active, Raman-active, and silent modes are denoted by I, R, and S, respectively.}
\setlength{\tabcolsep}{9pt}
\begin{tabular} {lcccccc}
\hline
\hline
Mode 		&	&$\nu$ &	&Eigen characters	\\
\hline
		&(meV)	&	&(cm$^{-1}$)   &	\\	
\hline
A$_2^{\prime\prime}$ (I)	&2.54	&	&20.5		&Hf+Ru+As	\\
E$^{\prime}$ (I+R)		&10.29	&	&83.0		&Hf+Ru+As	\\
E$^{\prime\prime}$ (R)		&10.82	&	&87.3		&Ru		\\
A$_1^{\prime}$ (R)		&14.52	&	&117.1		&Hf		\\
E$^{\prime\prime}$ (R)		&15.09	&	&121.7		&Hf		\\
A$_2^{\prime}$ (S)		&15.46	&	&124.7		&Hf+Ru		\\
E$^{\prime}$ (I+R)		&16.77	&	&135.3		&Hf+Ru+As	\\
E$^{\prime}$ (I+R)		&20.58	&	&166.0		&Ru+As	\\
A$_2^{\prime}$ (S)		&21.49	&	&173.3		&Ru		\\
A$_1^{\prime\prime}$ (S)	&25.68	&	&207.1		&As		\\
E$^{\prime}$ (I+R)		&27.85	&	&224.6		&Hf+Ru+As	\\
A$_2^{\prime\prime}$ (I)	&29.01	&	&234.0		&Hf+As		\\
A$_1^{\prime}$ (R)		&29.40	&	&237.1		&Ru		\\
E$^{\prime}$ (I+R)		&31.22	&	&251.8		&Ru+As	\\
E$^{\prime}$ (I+R)		&31.91	&	&257.4		&Ru+As		\\
A$_2^{\prime\prime}$ (I)	&33.36	&	&269.1		&Ru+As		\\
\hline
\hline
\end{tabular}
\end{table}

\section{Crystal structure and phonon dispersion}

The hexagonal phase of HfRuAs ($h$-HfRuAs) crystallizes in the Fe$_2$P-type structure with space group $P\bar{6}2m$ (No.~189) and point group $D_{3h}$, as illustrated in Fig.~1. The experimental lattice parameters are $a=b=6.568$~\AA \xspace and $c=3.842$~\AA \xspace~\cite{Meisner1983-3}. The crystal structure contains four inequivalent Wyckoff positions: Hf occupies the $3g$ ($x_{\mathrm{Hf}}$, 0, 1/2) site, Ru occupies the $3f$ ($x_{\mathrm{Ru}}$, 0, 0) site, As occupies the $2c$ (1/3, 2/3, 0) and $1b$ (0, 0, 1/2) sites, with internal structural parameters $x_{\mathrm{Hf}}=0.581$ and $x_{\mathrm{Ru}}=0.246$. 
The primitive unit cell of $h$-HfRuAs contains three formula units, corresponding to a total of nine atoms per unit cell. Consequently, the phonon spectrum consists of 27 vibrational branches, including three acoustic and twenty-four optical modes, as shown in Fig.~2(a). The absence of imaginary phonon frequencies throughout the BZ confirms the dynamical stability of the hexagonal phase.

The zone-center optical phonon modes can be classified according to the irreducible representations of the $D_{3h}$ point group as
$\Gamma(D_{3h}) = 2A_1^{\prime}+2A_2^{\prime}+6E^{\prime}+A_1^{\prime\prime}+3A_2^{\prime\prime}+2E^{\prime\prime}$.
Here, the $A$ and $E$ modes correspond to singly and doubly degenerate representations, respectively. The calculated phonon frequencies and corresponding vibrational characters are summarized in Table~I. Analysis of the phonon eigenvectors, schematically illustrated in Fig.~3, reveals that the atomic displacements associated with the $A_1^{\prime\prime}$, $A_2^{\prime\prime}$, and $E^{\prime\prime}$ modes are predominantly oriented along the out-of-plane ($z$) direction. In contrast, the displacements corresponding to the $A_1^{\prime}$, $A_2^{\prime}$, and $E^{\prime}$ modes are mainly confined within the in-plane ($xy$) directions.

The calculated phonon dispersion relations together with the mode-resolved EPC strength are shown in Fig.~2(a). The red circles superimposed on the phonon branches represent the magnitude of the EPC strength. The largest EPC contributions are concentrated in the low-frequency acoustic and low-lying optical phonon branches below approximately 10~meV, particularly around the $\Gamma$ point. In contrast, the higher-frequency optical phonon modes exhibit comparatively weaker EPC contributions. This behavior indicates that the superconductivity in $h$-HfRuAs is predominantly governed by low-energy lattice vibrations. The calculated phonon density of states (PhDOS), shown in Fig.~2(b), reveals distinct atomic contributions across different frequency regions. The modes upto 16~meV are primarily associated with Hf atoms with secondary contributions from Ru atoms. Notably, this frequency region contributes approximately 82\% of the total EPC. In the frequency window of 16--25~meV, the phonon modes are predominantly governed by Ru vibrations. The modes above 25~meV are mainly associated with As vibrations.  

The calculated Eliashberg spectral function $\alpha^2F(\omega)$ and cumulative EPC constant $\lambda(\omega)$ are presented in Fig.~2(c). Pronounced peaks in $\alpha^2F(\omega)$ are observed below 10~meV, indicating that the low-frequency phonon modes provide the dominant contribution to the EPC. These low-energy phonons contribute nearly 55\% of the total EPC strength and mainly originate from the vibrations dominated by Hf and Ru atoms. The cumulative EPC constant rapidly increases within this low-frequency region and gradually saturates at higher frequencies, reaching a total EPC value of $\lambda \approx 1.56$. Such a large EPC strength places $h$-HfRuAs in the strong-coupling regime and demonstrates that superconductivity in this compound is predominantly driven by strong coupling between electrons and low-frequency phonon modes.


\begin{figure}
\includegraphics[width=80mm]{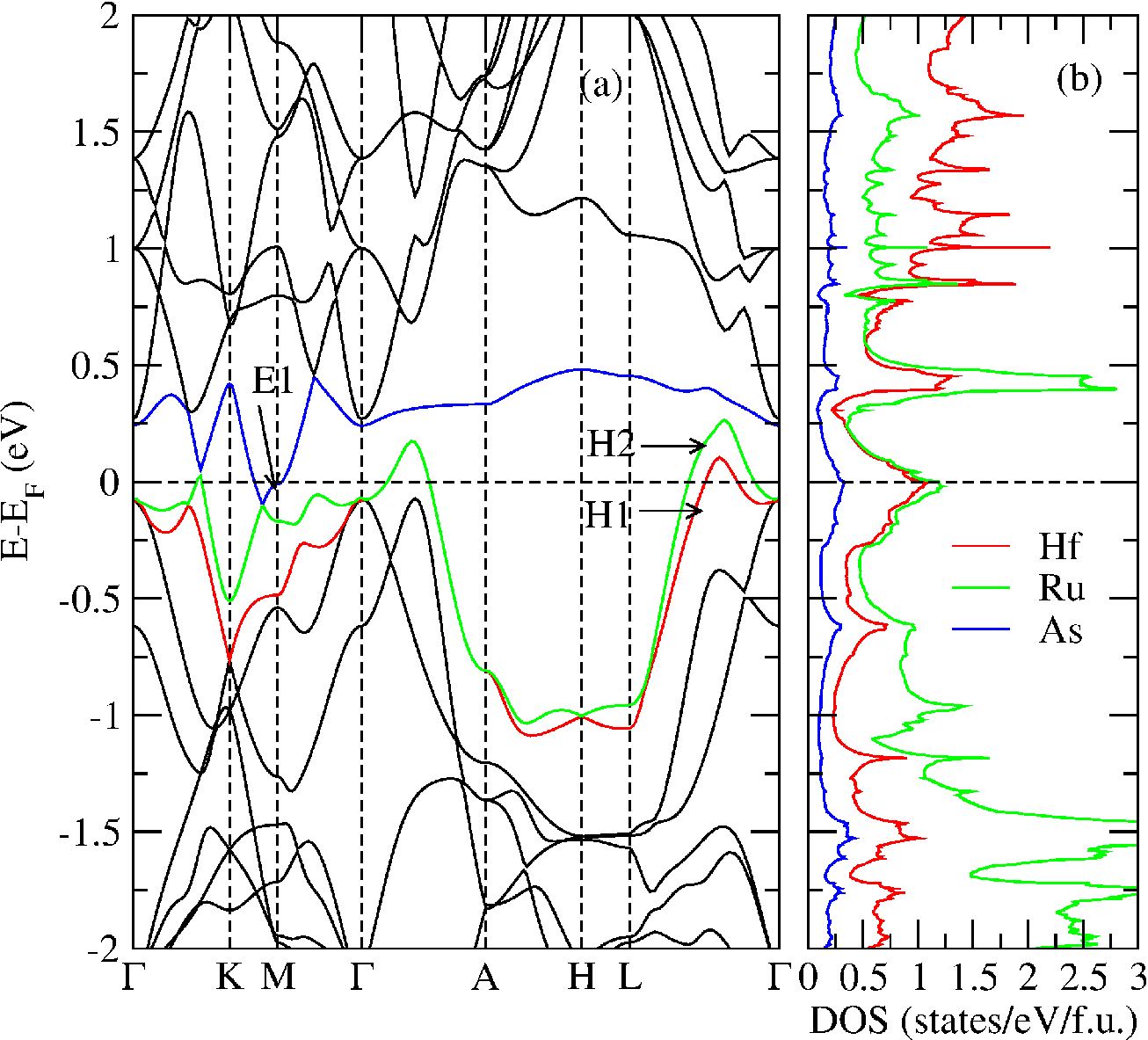}
\caption{(a)Band structure of $h$-HfRuAs. Three bands which cross the $E_F$ are indicated with red, green and blue colours. The H1 and H2 indicates the hole nature and E1 indicates the electron nature of the Fermi surface. (b) Atom projected DOSs of $h$-HfRuAs.}
\end{figure}

\begin{figure}
\includegraphics[width=80mm]{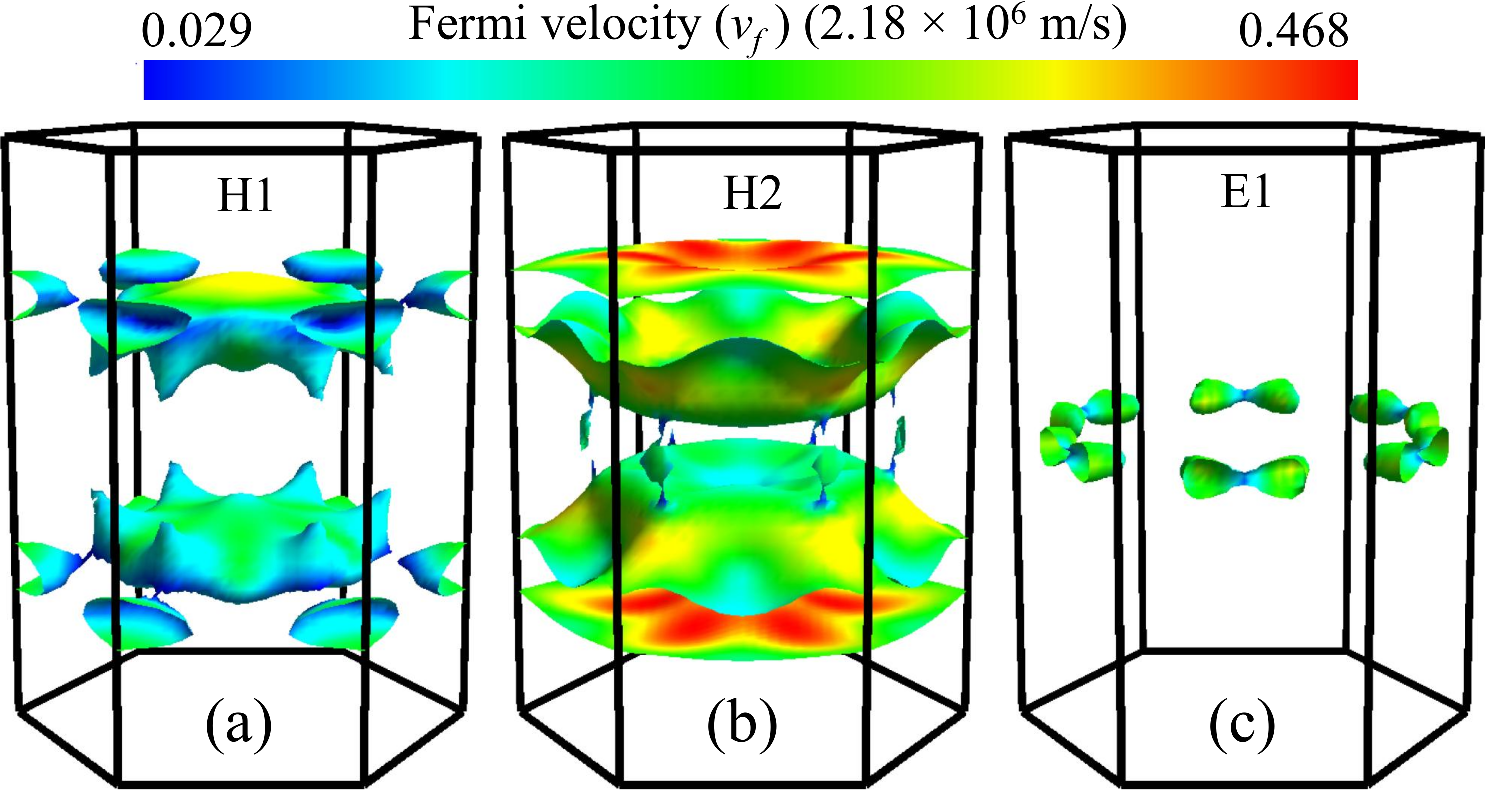}
\caption{Momentum {\bf k}-dependent magnitude of the Fermi velocity ($v_f$) on the FS of $h$-HfRuAs on (a) hole pocket H1, (b) H2, and (c) electron pocket E1.}
\end{figure}

\section{Electronic structure and Fermi surface topology}

The calculated electronic band structure of $h$-HfRuAs is shown in Fig.~4(a). Three bands cross the $E_F$, labeled as H1, H2, and E1. The H1 and H2 bands intersect the $E_F$ mainly along the $\Gamma$--K, $\Gamma$--A, and L--$\Gamma$ directions, forming hole-like FS sheets, as illustrated in Figs.~5(a) and 5(b). In contrast, the E1 band crosses the $E_F$ primarily along the K--M direction, giving rise to an electron-like FS pocket [Fig.~5(c)]. The coexistence of both hole- and electron-like FS sheets indicates a multiband electronic structure near $E_F$, which is favorable for enhanced interband electron--phonon scattering and SC pairing interactions.

The FS pockets associated with the H1 and H2 bands exhibit complex multi-sheet topology with pronounced anisotropic features, whereas the E1 band forms a comparatively simple electron pocket. In particular, the H2 FS displays strong momentum-dependent anisotropy, which later manifests in the anisotropic EPC strength and SC gap distribution. The calculated momentum-resolved Fermi velocity ($v_f$) further reveals significant anisotropy across different FS sheets. Large $v_f$ regions on the H2 sheet originate from highly dispersive electronic bands and correspond to relatively light effective masses, indicating strong electronic sensitivity to lattice vibrations. Such anisotropic electronic dispersion can substantially influence the momentum dependence of the electron--phonon matrix elements and generate direction-dependent EPC strengths over the FS. Conversely, regions with lower $v_f$ can enhance the electronic residence time and favor Cooper pairing.

The calculated electronic density of states (DOS) is presented in Fig.~4(b). The atom-projected DOS at the $E_F$ is approximately 1.07, 1.18, and 0.65 states/eV/atom for Hf, Ru, and As atoms, respectively, resulting in a total DOS of about 2.9 states/eV/f.u at $E_F$. The electronic states near the $E_F$ are predominantly contributed by the Hf-$d$ and Ru-$d$ orbitals together with smaller contributions from the As-$p$ orbitals, indicating significant hybridization between transition-metal $d$ states and pnictogen $p$ states. Such orbital hybridization enhances the electron--phonon matrix elements and plays an important role in strengthening the EPC, consistent with the strong-coupling SC behavior discussed in the following sections.

\begin{figure}
\includegraphics[width=80mm]{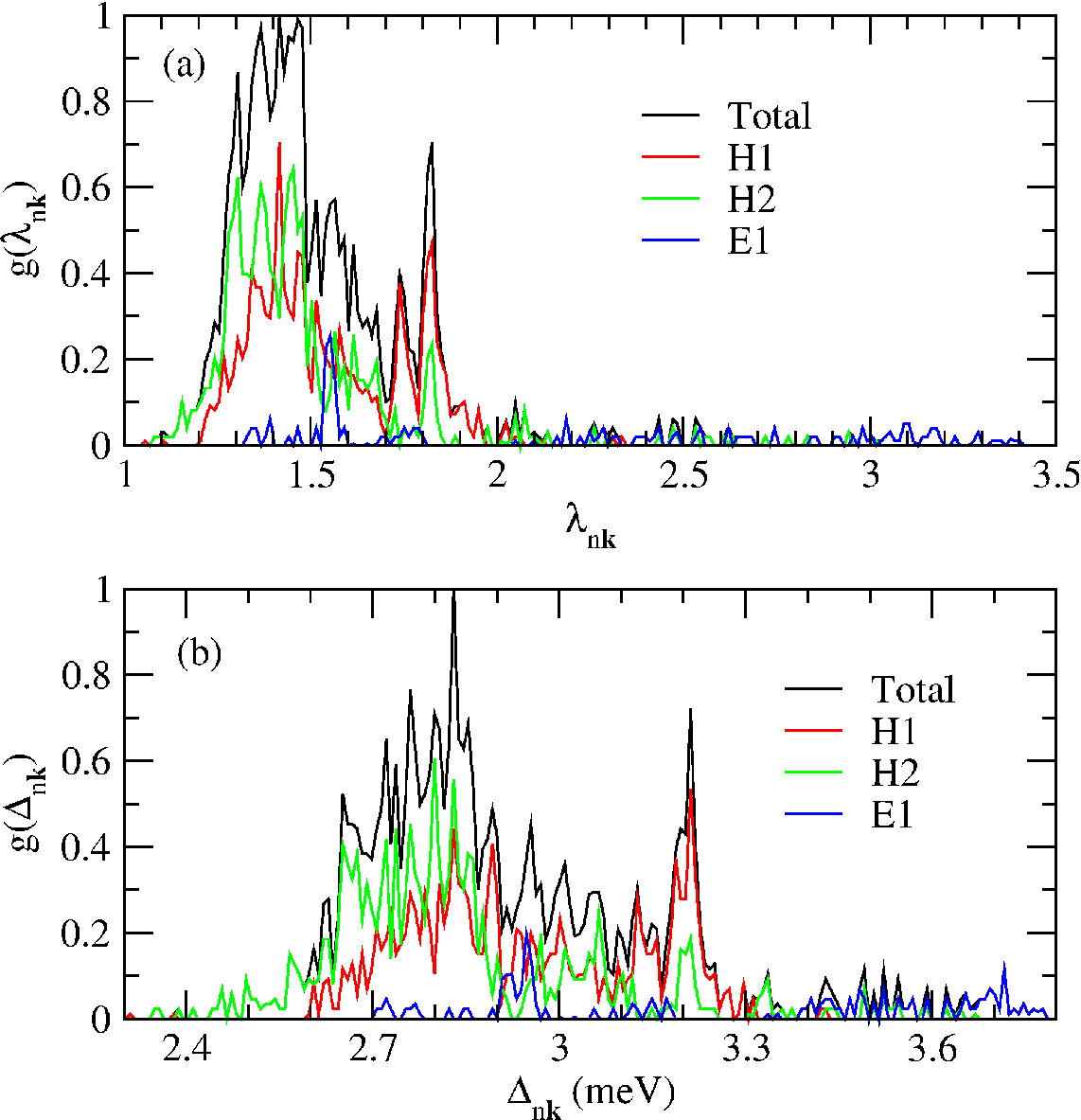}
\caption{Total and band-resolved normalized density distribution of (a) the EPC strength $\lambda_{n\mathbf{k}}$, g($\lambda_{n\mathbf{k}}$), and of (b) SC gap $\Delta_{n\mathbf{k}}$ and g($\Delta_{n\mathbf{k}}$) at the temperature 0.5 K on the FS of $h$-HfRuAs.}
\end{figure}

\begin{figure}
\includegraphics[width=80mm]{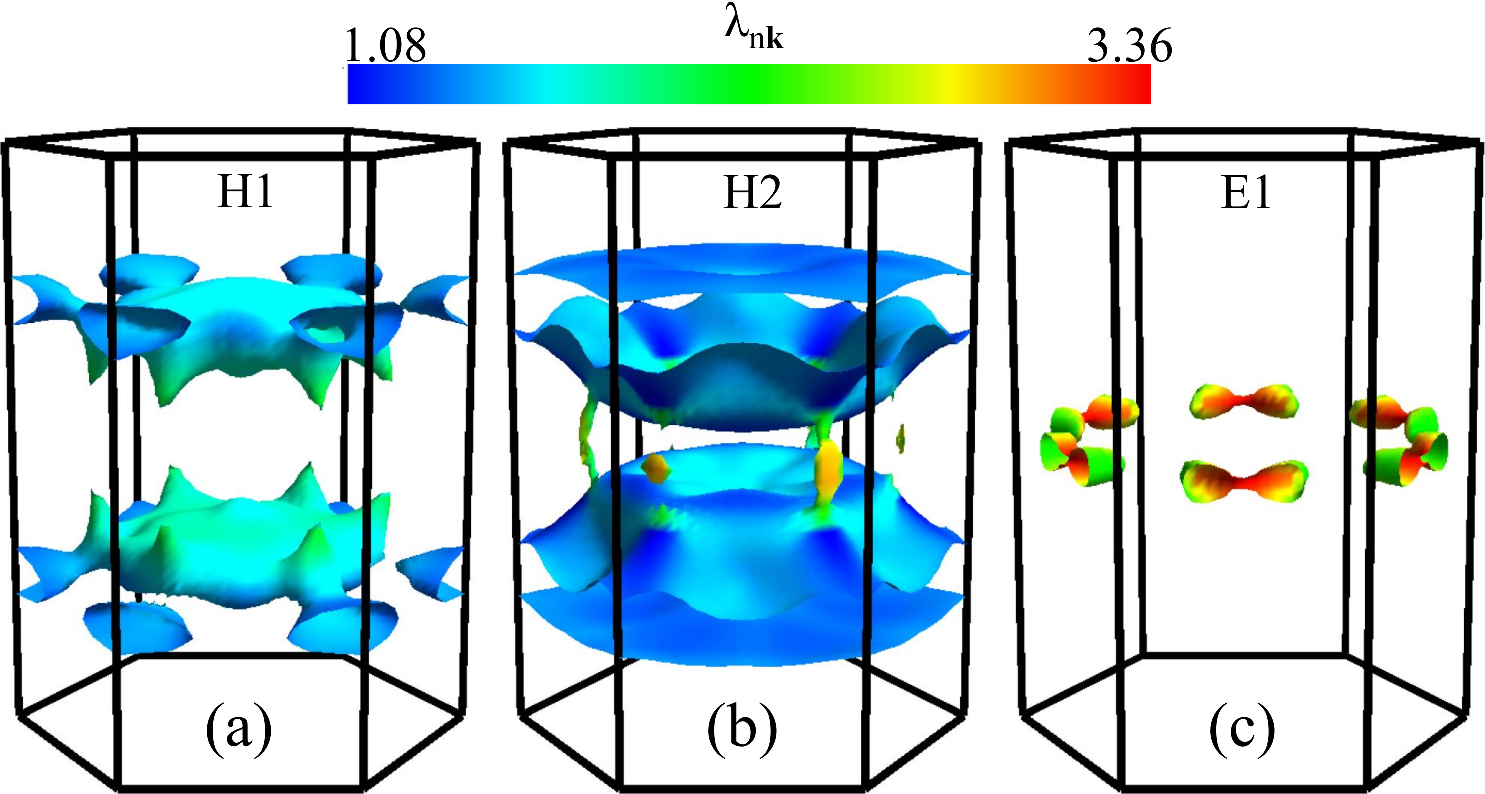}
\caption{Momentum {\bf k}-dependent EPC strength $\lambda_{n\mathbf{k}}$ on the FS of $h$-HfRuAs on (a) hole pocket H1, (b) hole pocket H2, and (c) electron pocket E1.}
\end{figure}

\section{Electron--Phonon Coupling}

For conventional superconductors driven by electron--phonon interactions, the ME framework provides a reliable description of both the SC transition temperature ($T_c$) and the anisotropic SC gap structure. Within this formalism, two key quantities are required: the Eliashberg spectral function $\alpha^2F(\omega)$ and the effective screened Coulomb repulsion parameter $\mu_c^*$~\cite{Margine2013}. The Coulomb interaction enters the anisotropic ME equations through the momentum-dependent matrix elements $V(\mathbf{k}-\mathbf{k}')$~\cite{Lee2023}. By performing a double average of this interaction over the FS, one obtains an effective Coulomb parameter expressed as
$\mu_c = N_F \langle\!\langle V(\mathbf{k}-\mathbf{k}') \rangle\!\rangle_{\mathrm{FS}}$,
where $N_F$ is the electronic density of states at the $E_F$. As mentioned in Section II, in practical implementations of the ME formalism, this quantity is replaced by $\mu_c^*$, which effectively incorporates retardation effects arising from the separation between electronic and phononic energy scales.

In previous theoretical studies~\cite{Reddy2014,Reddy2016,Reddy2017,Reddy2017_1,Sen2020,Babu2023,Keshri2025}, $\mu_c^*$ is commonly treated as an adjustable parameter~\cite{Ponce2016,Lee2023,Noffsinger2010}. For conventional elemental superconductors, a typical choice is $\mu_c^* \approx 0.10$~\cite{Pellegrini2022}. More generally, values in the range of 0.05--0.10 are usually appropriate for weakly correlated $sp$-bonded systems, whereas larger values of 0.10--0.15 are commonly employed for transition-metal-based compounds~\cite{Pellegrini2022,Allen1983,Flores-Livas2020}. In certain strongly correlated or transition-metal systems, even larger values of $\mu_c^* > 0.2$ may be required to reproduce experimental SC properties~\cite{Giustino2007,Liu1996,Bauer2012}. Since the electronic states near the $E_F$ in $h$-HfRuAs are dominated by Hf-$d$ and Ru-$d$ orbitals (see Appendix A and Fig. 10), a value of $\mu_c^*=0.15$ was adopted in the present calculations. The effect of the $\mu_c^*$ value on the calculated $T_c$ and SC gap function is discussed in Appendix B. 

The Eliashberg spectral function $\alpha^2F(\omega)$, shown in Fig.~2(c), provides a frequency-resolved description of the EPC. The low-frequency region is characterized by several pronounced peaks arising from the combined contributions of acoustic branches and low-lying optical phonon modes. These phonons extend up to approximately 10~meV in the phonon dispersion and predominantly involve coupled vibrations of Hf and Ru atoms. Since low-energy phonons contribute more strongly to the EPC through the $1/\omega$ weighting factor, these modes dominate the SC pairing interaction.

The total EPC constant $\lambda$, often referred to as the mass-enhancement parameter, is obtained by integrating $\alpha^2F(\omega)/\omega$ over the full phonon spectrum. For $h$-HfRuAs, the calculated EPC strength $\lambda = 1.56$ indicates strong EPC. This value exceeds those reported for closely related compounds, including h-ZrRuP ($\lambda = 1.25$~\cite{Bagci2019} and 1.1~\cite{Salamati1997}), h-ZrRuAs ($\lambda = 1.32$~\cite{Tutuncu2020} and 0.79~\cite{Ren2025}), and h-HfRuP ($\lambda = 0.87$~\cite{Ren2025}). The relatively large EPC places $h$-HfRuAs within the class of strong-coupling superconductors, comparable to elemental superconductors such as Pb and Hg~\cite{PBAllen1975}, A15 compounds including Nb$_3$Sn and Nb$_3$Ge~\cite{Stewart2015,Semenok2025}, as well as high-pressure hydrides such as H$_3$S~\cite{Drozdov2015}.

Figure~6(a) presents the normalized distributions of the total and band-resolved momentum-dependent EPC strength $\lambda_{n\mathbf{k}}$ projected onto the FS. Here, $\lambda_{n\mathbf{k}}$ characterizes the momentum-resolved EPC and provides a direct measure of the coupling strength associated with each electronic state on the FS. The total $\lambda_{n\mathbf{k}}$ distribution is centered around $\lambda_{n\mathbf{k}} \approx 1.56$ and exhibits a substantial spread, reflecting pronounced anisotropy of the EPC over different FS sheets. The band-resolved decomposition further reveals that the dominant contribution to the total EPC originates from the H1 and H2 hole-like bands, whereas the E1 electron pocket contributes comparatively less to the overall coupling strength.

The momentum-resolved EPC projected onto the individual FS sheets is shown in Fig.~7. Significant anisotropy is clearly observed across different regions of the FS, particularly on the H2 sheet, where $\lambda_{n\mathbf{k}}$ exhibits large spatial variations. Since $\lambda_{n\mathbf{k}}$ directly enters the quasiparticle mass-renormalization factor through the relation
$Z_{n\mathbf{k}} = 1+\lambda_{n\mathbf{k}}$,
strong EPC leads to substantial renormalization of the electronic quasiparticle properties. In particular, the $v_f$ is reduced according to $v_{n\mathbf{k}} \rightarrow v_{n\mathbf{k}}/Z_{n\mathbf{k}}$, while the effective mass and electronic DOS near the $E_F$ are correspondingly enhanced. Within a simple parabolic-band approximation, the effective-mass enhancement approximately follows $m^*/m \sim Z_{n\mathbf{k}}$.

Such EPC induced renormalization effects can be experimentally probed through thermodynamic measurements, especially via the electronic specific heat. In the low-temperature limit, the electronic specific heat is expressed as $C_e = \gamma_n T$, where the Sommerfeld coefficient is given by
$\gamma_n = \frac{\pi^2}{3} k_B^2 N_F = 2.359 \times N_F \;\mathrm{mJ/mol\,K^2}$.
Using the calculated DOS value of $N_F = 2.9$ states/eV/f.u. for $h$-HfRuAs, the bare Sommerfeld coefficient is estimated to be $\gamma_n \approx 6.8$~mJ/mol\,K$^2$. Inclusion of EPC renormalization enhances this value by a factor of $(1+\lambda)$, yielding a renormalized Sommerfeld coefficient of approximately $\gamma_n \approx 17.5$~mJ/mol\,K$^2$ for $\lambda = 1.56$.

Experimentally, a Sommerfeld coefficient of $\gamma = 8.88$~mJ/mol\,K$^2$ has been reported for $h$-HfRuAs~\cite{Zhou2025}. Although this experimental value indicates moderate quasiparticle mass enhancement relative to the bare electronic DOS, it remains substantially smaller than the fully renormalized theoretical estimate. This discrepancy may originate from extrinsic effects such as phase coexistence, disorder, impurity scattering, or deviations from ideal stoichiometry in experimental samples, whereas the present calculations represent the intrinsic EPC strength of the ideal hexagonal phase.

\begin{figure}
\includegraphics[width=80mm]{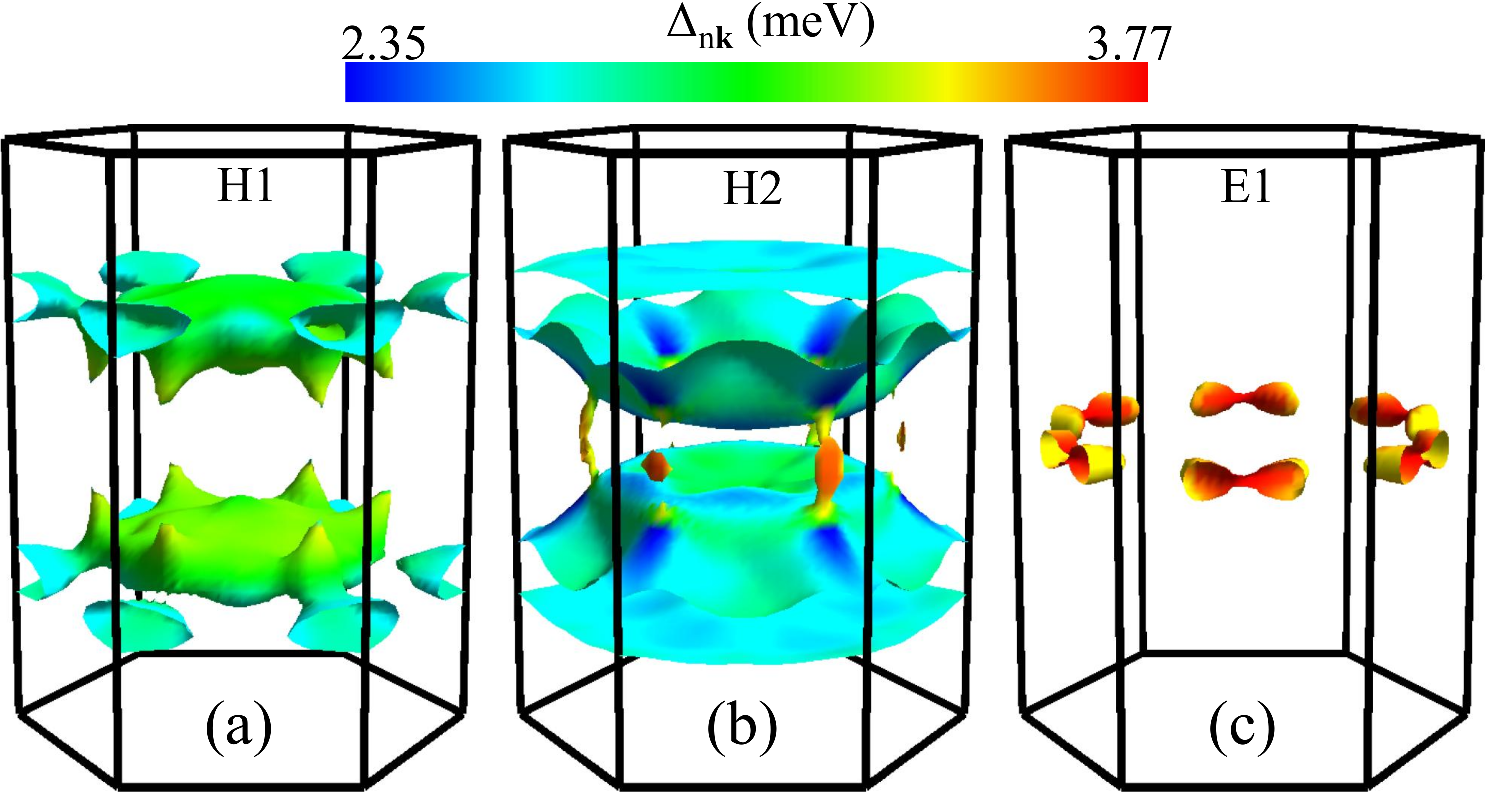}
\caption{Electronic state {\bf k}-dependent SC gap  $\Delta_{n\mathbf{k}}$ on the FS of $h$-HfRuAs at the temperature 0.5 K on (a) hole pocket H1, (b) H2, and (c) electron pocket E1.}
\end{figure}

\begin{figure}
\includegraphics[width=80mm]{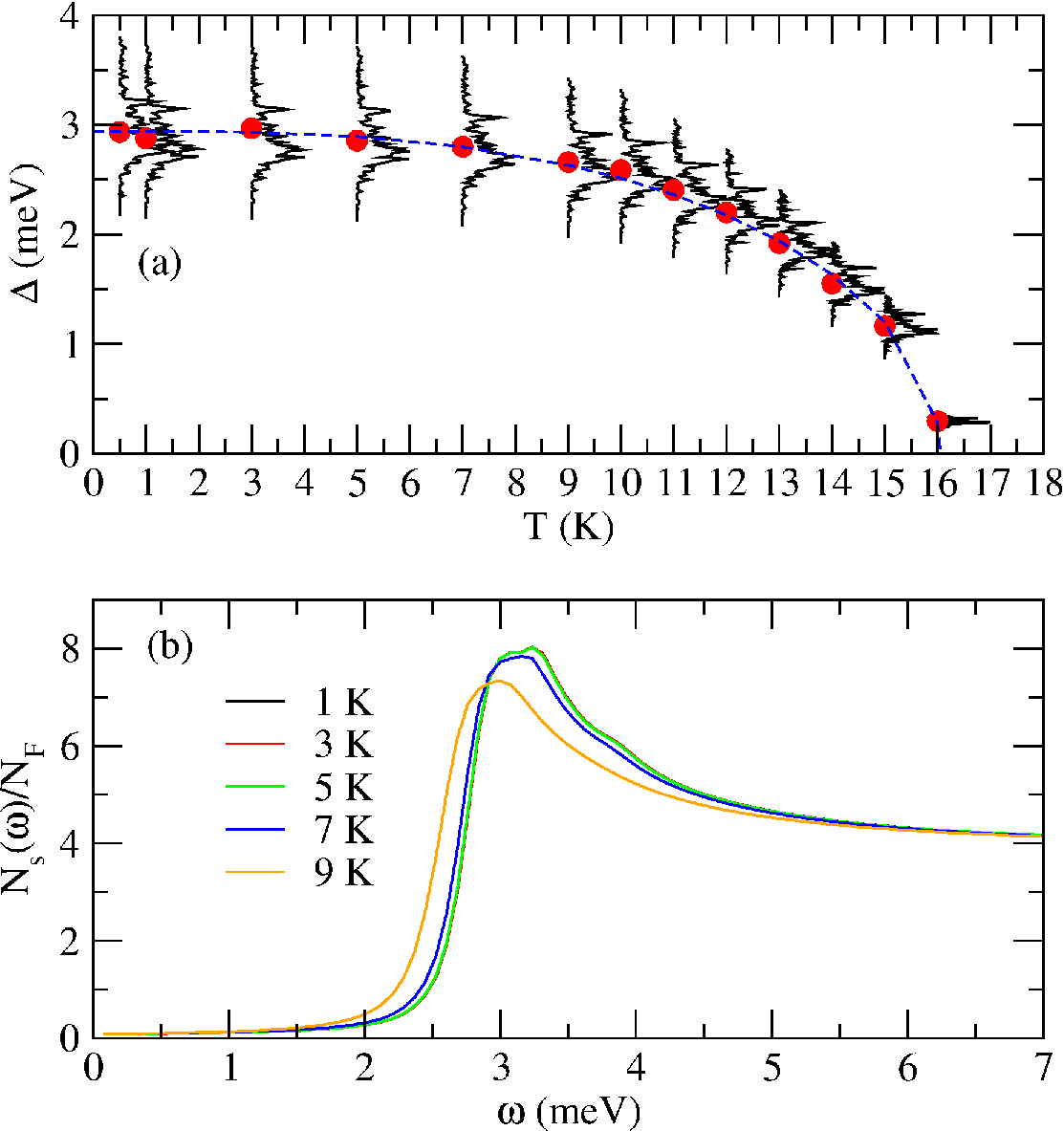}
\caption{(a) SC gap $\Delta$ as a function of temperature T , with the red dots denoting the averaged gap values, which are fitted using a BCS-type temperature dependence [Eq. (9)] (the dashed blue curve). (b) Normalized SC quasiparticle DOS N$_s(\omega)$/N$_F$ as a function of energy calculated at several temperatures.}
\end{figure}

\section{Superconducting Properties}

To determine the SC gap function $\Delta_{n\mathbf{k}}$, the fully anisotropic ME equations were solved self-consistently at several temperatures using a Coulomb pseudopotential $\mu_c^* = 0.15$. This choice is motivated by the fact that the electronic states near the $E_F$ are predominantly derived from the Hf-$d$ and Ru-$d$ orbitals (see Fig.~10), for which relatively large values of $\mu_c^*$ are generally appropriate in transition-metal-based superconductors.

Figure~6(b) presents the normalized distributions of the total and band-resolved SC gap $\Delta_{n\mathbf{k}}$ evaluated at $T = 0.5$~K. The distribution is centered around $\Delta_{n\mathbf{k}} \approx 2.9$~meV and exhibits a substantial spread of approximately 0.8~meV, indicating pronounced anisotropy of the SC gap over the FS. The broad distribution of $\Delta_{n\mathbf{k}}$ closely follows the anisotropic distribution of the momentum-resolved EPC strength $\lambda_{n\mathbf{k}}$, reflecting the strong coupling between the SC pairing interaction and the anisotropic EPC.

The momentum-resolved SC gap projected onto the FS at $T = 0.5$~K is shown in Fig.~8. A finite SC gap opens on all three FS sheets, confirming a fully gapped SC state. Nevertheless, the magnitude of $\Delta_{n\mathbf{k}}$ varies significantly across different regions of the FS. Among the three sheets, the hole-like H2 band exhibits the strongest gap anisotropy [Fig.~8(b)], whereas the electron-like E1 pocket shows comparatively weaker variation [Fig.~8(c)]. The SC gap ranges from a minimum value of approximately 2.35~meV to a maximum value of approximately 3.77~meV, both occurring on the H2 sheet. The largest gap values are located along the $\Gamma$--K direction near the K point, particularly on the connecting regions linking the upper and lower portions of the H2 FS. Enhanced gap values are also observed near the M point along the $\Gamma$--M direction on the disconnected truncated dumbbell-like pockets of the E1 sheet.

The SC gap distributions shown in Figs.~6(b) and 8 indicate that $h$-HfRuAs is characterized by a single anisotropic SC gap rather than distinct multiple gaps associated with different FS sheets. Despite the pronounced momentum dependence of $\Delta_{n\mathbf{k}}$, the SC state remains fully gapped and consistent with conventional spin-singlet $s$-wave pairing symmetry. A clear correspondence between the distributions of $\lambda_{n\mathbf{k}}$ and $\Delta_{n\mathbf{k}}$ is evident from Figs.~6--8, indicating that regions with stronger EPC generally develop larger SC gaps. Such behavior is expected for a phonon-mediated superconductor within the anisotropic ME formalism.

The anisotropy of the SC gap can be further understood from the orbital character of the electronic states near the $E_F$ shown in Fig.~10. The SC gaps on the H1 and H2 FS sheets predominantly originate from hybridized Hf-$d$ and Ru-$d$ orbitals, whereas the E1 pocket mainly derives from the $d_{zx}$ orbitals of both Hf and Ru atoms. The orbital-dependent electronic structure therefore plays an important role in generating the anisotropic EPC and the corresponding variation of the SC gap over the FS.

The temperature evolution of the SC gap is shown in Fig.~9(a). To estimate the SC $T_c$, the averaged SC gap values at different temperatures [red dots in Fig.~9(a)] were fitted using a phenomenological BCS-type temperature dependence~\cite{Johnston2013},
\begin{equation}
\Delta(T) = \Delta(0)\tanh \left[ \alpha\sqrt{\frac{T_c}{T}-1}\right].
\end{equation}
The fitting yields $\Delta(0)=2.94$~meV, $\alpha = 1.64$, and $T_c$ = 16.06~K. The calculated $T_c$ is significantly higher than the experimentally reported values of 4.37--4.93~K~\cite{Meisner1983-1} and 7.25~K~\cite{Zhou2025} for $h$-HfRuAs. The obtained value of $\alpha = 1.64$ is slightly smaller than the weak-coupling BCS value of 1.73. This deviation can be attributed to the fact that the empirical BCS-type expression assumes an isotropic single-band SC gap, whereas $h$-HfRuAs exhibits a multiband SC state with pronounced momentum-dependent gap anisotropy. Consequently, the fitted $\alpha$ represents an effective average over anisotropic SC gaps distributed across multiple FS sheets.

Using the obtained values of $\Delta(0)$ and $T_c$, the SC gap ratio is estimated to be $2\Delta(0)/k_BT_c \approx 4.25$, which significantly exceeds the weak-coupling BCS limit of 3.53~\cite{Bardeen1957,Johnston2013}. This enhancement, together with the large EPC constant $\lambda \approx 1.56$, provides strong evidence that superconductivity in $h$-HfRuAs belongs to the strong-coupling regime.

Using the calculated Fermi velocity $v_f$ [Fig.~5] and SC gap, the coherence length $\xi_0$ can be estimated within the BCS approximation using the relation
$\xi_0=\frac{\hbar v_f}{\pi\Delta}$.
Using the averaged values $v_f \approx 0.5 \times 10^6$~m/s and $\Delta \approx 2.94$~meV, the estimated coherence length is $\xi_0 \approx 35$~nm. This value is approximately 6.8 times larger than the experimentally reported Ginzburg--Landau coherence length $\xi_{GL}\sim5.15$~nm~\cite{Zhou2025}. Such a discrepancy is expected because the BCS expression corresponds to the intrinsic clean-limit coherence length, whereas the experimentally extracted $\xi_{GL}$ is obtained from the upper critical field and reflects the dirty-limit regime. In polycrystalline samples, structural disorder, phase coexistence, and impurity scattering substantially reduce the electronic mean free path $l$, leading to a suppressed effective coherence length approximately following $\xi_{GL}\sim\sqrt{\xi_0 l}$. Furthermore, the experimentally determined $\xi_{GL}$ additionally incorporates effects associated with vortex physics and sample inhomogeneity, and therefore cannot be directly compared with the clean-limit BCS estimate.

The normalized quasiparticle density of states $N_s(\omega)/N_F$, calculated using Eq.~(8), is presented in Fig.~9(b) at several temperatures. The spectra exhibit a well-defined SC gap accompanied by a single coherence peak at the gap edge, indicating the absence of multiple distinct gap features and thereby supporting a single anisotropic gap SC scenario in $h$-HfRuAs. Furthermore, the low-energy region exhibits a fully opened U-shaped profile, reflecting the absence of nodes in the SC order parameter. Such behavior is characteristic of a conventional spin-singlet $s$-wave SC state, consistent with previous experimental and theoretical studies~\cite{Keshri2025,Zheng2017}.

\begin{figure}
\includegraphics[width=80mm]{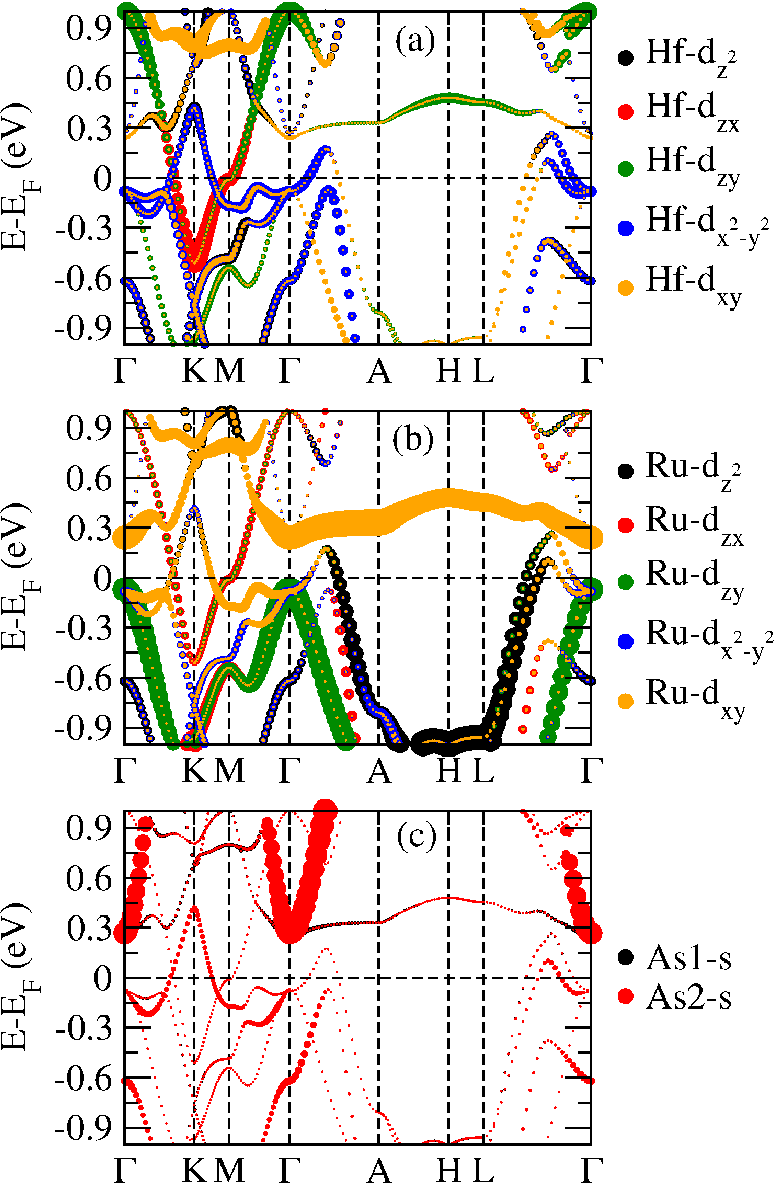}
\caption{Orbital-projected electronic band structure of $h$-HfRuAs showing contributions from (a) Hf-$d$, (b) Ru-$d$, and (c) As-$s$ orbitals.}
\end{figure}

\begin{figure}
\includegraphics[width=80mm]{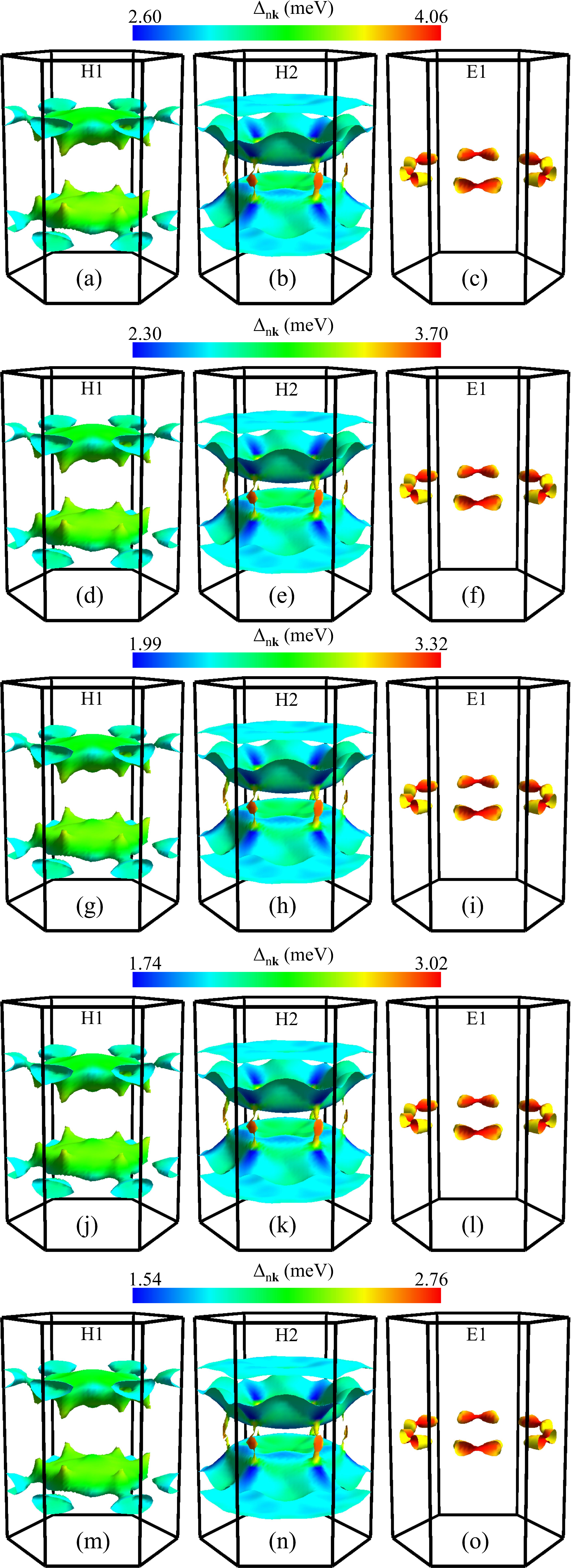}
\caption{Momentum {\bf k}-dependent SC gap  $\Delta_{n\mathbf{k}}$ on the FS of $h$-HfRuAs at the temperature 3.0 K using $\mu_c^*$ of (a)-(c) 0.11, (d)-(f) 0.15, (g)-(i) 0.20, (j)-(l) 0.25, and (m)-(o) 0.30.}
\end{figure}

\section{Conclusions}

In this work, we have investigated the SC properties of $h$-HfRuAs using fully anisotropic ME calculations combined with first-principles electron--phonon Wannier interpolation. Our results establish $h$-HfRuAs as a phonon-mediated strong-coupling superconductor with a sizeable electron--phonon coupling strength of $\lambda \approx 1.56$. The superconductivity is found to originate predominantly from low-frequency phonon modes associated mainly with Hf and Ru vibrations. The momentum-resolved analysis reveals pronounced anisotropy in both the EPC and SC gap across different FS sheets, particularly on the hole-like H2 pocket. Nevertheless, the SC state remains fully gapped and is characterized by a single anisotropic spin-singlet $s$-wave gap without nodal features. The calculated SC gap magnitude and enhanced gap ratio $2\Delta(0)/k_B T_c$ indicate clear deviations from weak-coupling BCS behavior and confirm the strong-coupling nature of superconductivity in this system. The substantial quasiparticle renormalization and enhanced Sommerfeld coefficient further support this conclusion. Although some quantitative differences remain between the present theoretical predictions and currently available experimental measurements, our study provides a consistent microscopic description of superconductivity in $h$-HfRuAs and highlights the important role of momentum-dependent electron--phonon interactions in multiband superconductors. These findings not only advance the understanding of superconductivity in the TT$^{\prime}$X family of compounds, but also provide useful guidance for future experimental investigations on high-quality single crystals.

\begin{acknowledgments}
The authors acknowledge support from the National Science and Technology Council (NSTC) and the National Center for Theoretical Sciences (NCTS), Taiwan. They also gratefully acknowledge the National Center for High-Performance Computing (NCHC), Taiwan, for providing computational resources.
\end{acknowledgments}

\appendix

\section{Orbital projected band structure}
The orbital-projected electronic band structure of $h$-HfRuAs is shown in Fig.~10. The electronic states near the Fermi level ($E_F$) are predominantly contributed by the Hf-$d_{zx}$, Hf-$d_{x^2-y^2}$, and Hf-$d_{xy}$ orbitals together with the Ru-$d_{z^2}$, Ru-$d_{zx}$, and Ru-$d_{xy}$ orbitals. In addition, comparatively smaller contributions from the As-$s$ states are also present near $E_F$. The strong hybridization between Hf-$d$ and Ru-$d$ orbitals plays an important role in shaping the multiband FS topology and enhancing the EPC strength as discussed in the main text.

\begin{table}
\caption{SC transition temperature $T_c$ and averaged SC gap $\Delta_{\mathrm{avg}}$ on the FS for different values of the Coulomb pseudopotential $\mu_c^*$. The transition temperatures are estimated using the Allen--Dynes modified McMillan (ADM) formula [Eq.~(B1)] with the calculated EPC constant $\lambda = 1.56$ and logarithmic phonon frequency $\hbar\omega_{\log}=10.4125$ meV.}
\setlength{\tabcolsep}{8pt}
\begin{tabular}{lcccccc}
\hline
\hline
$\mu_c^*$ & 0.11 & 0.15 & 0.20 & 0.25 & 0.30 \\
\hline
$T_c$ (K) & 13.9 & 12.3 & 10.3 & 8.3 & 6.5 \\
$\Delta_{\mathrm{avg}}$ (meV) &3.20 &2.97 &2.53 &2.26 &2.03 \\
\hline
\hline
\end{tabular}
\end{table}

\section{Effect of Coulomb Pseudopotential $\mu_c^*$ on $T_c$ and the SC Gap Function}

As discussed in the main text, the $\mu_c^*$ is an empirical parameter and its value affects the calculated SC $T_c$ and SC gap function $\Delta_{n\mathbf{k}}$. In general, increasing $\mu_c^*$ suppresses superconductivity, leading to lower $T_c$ values and reduced gap magnitudes. A rigorous evaluation of the $\mu_c^*$ dependence of $T_c$ requires solving the fully anisotropic ME equations [Eqs.~(1) and (2)] over a broad temperature range until the SC gap vanishes. However, such calculations are computationally demanding, particularly for systems with relatively small SC gaps~\cite{Keshri2025} and in the temperature regime close to $T_c$, where numerical convergence becomes increasingly difficult.

To investigate the evolution of $T_c$ with $\mu_c^*$ in $h$-HfRuAs, we employed the simplified Allen--Dynes modified McMillan expression~\cite{Lee2023}, as introduced in Eq.~(B1), 
\begin{equation}
k_B T_c=\frac{\hbar\omega_{\log}}{1.20}\exp\left[\frac{-1.04(1+\lambda)}{\lambda-\mu_c^{*}(1+0.62\lambda)}\right].
\label{eq:B1}
\end{equation}
together with the calculated EPC constant $\lambda$ and logarithmic phonon frequency $\hbar\omega_{\log}$. The resulting $T_c$ values are summarized in Table~II. In addition, to explicitly examine the influence of $\mu_c^*$ on the SC gap structure, the anisotropic ME equations were solved at $T = 3.0$~K for five representative values of $\mu_c^*$. The corresponding momentum-resolved SC gaps $\Delta_{n\mathbf{k}}$ are presented in Fig.~11, while the averaged gap values $\Delta_{\mathrm{avg}}$ are listed in Table~II.

The calculated results reveal a systematic suppression of superconductivity with increasing $\mu_c^*$. Specifically, increasing $\mu_c^*$ from 0.15 to 0.30 lowers the estimated $T_c$ from 12.3~K to 6.5~K, corresponding to a reduction of approximately 5.8~K. We further observe that the $T_c$ obtained directly from the anisotropic ME calculations are consistently higher than those predicted by the simplified Allen--Dynes expression. For example, at $\mu_c^* = 0.15$, the anisotropic ME calculations yield $T_c \approx 16$~K, whereas the Allen--Dynes estimate gives $T_c$ $\approx$ 12.3~K.

Interestingly, although the overall magnitude of the SC gap decreases with increasing $\mu_c^*$, the momentum-space distribution and anisotropic character of $\Delta_{n\mathbf{k}}$ remain nearly unchanged, as illustrated in Fig.~11. This robustness indicates that the essential anisotropic gap features are largely insensitive to the variations in the Coulomb pseudopotential. Consequently, the primary conclusions of the present work---namely the strong electron--phonon interaction and anisotropic SC gap behavior in $h$-HfRuAs---are not qualitatively affected by the particular choice of $\mu_c^*$.

We note, however, that treating $\mu_c^*$ as an adjustable semiempirical parameter remains a limitation of conventional ME-based approaches. Recent theoretical developments have enabled fully \textit{ab initio} descriptions of superconductivity in which the effective electron--electron Coulomb interaction is evaluated directly from first principles within an extended ME framework~\cite{Pellegrini2022}. Nevertheless, performing such fully parameter-free calculations for the present system is computationally beyond the scope of the present study.







\end{document}